\documentclass[aps,twocolumn]{revtex4}
\usepackage{epsf}

\begin{document}

\title{A Coupled Oscillator Model for
       Grover's Quantum Database Search Algorithm%
\footnote{Winning project in physics category at the National Science Fair
          IRIS 2006 (Initiative for Research and Innovation in Science
          promoted by Intel-DST-CII), New Delhi, India, December 2006.
          Presented at the Intel ISEF 2007 (International Science and
          Engineering Fair), Albuquerque, USA, May 2007.}}
\author{Aavishkar A. Patel}
\email{teslawolf@gmail.com}
\affiliation{National Public School, Rajajinagar,
             Bangalore-560010, India}
\author{{\sl Guide:} Apoorva D. Patel}
\email{adpatel@cts.iisc.ernet.in}
\affiliation{Centre for High Energy Physics, Indian Institute of Science,
             Bangalore-560012, India}

\begin{abstract}
Grover's database search algorithm is the optimal algorithm for finding
a desired object from an unsorted collection of items. Although it was
discovered in the context of quantum computation, it is simple and versatile
enough to be implemented using any physical system that allows superposition
of states, and several proposals have been made in the literature. I study
a mechanical realisation of the algorithm using coupled simple harmonic
oscillators, and construct its physical model for the simplest case of four
identical oscillators. The identification oracle is implemented as an elastic
reflection of the desired oscillator, and the overrelaxation operation is
realised as evolution of the system by half an oscillation period. I derive
the equations of motion, and solve them both analytically and by computer
simulation. I extend the ideal case analysis and explore the sensitivity of
the algorithm to changes in the initial conditions, masses of springs and
damping. The amplitude amplification provided by the algorithm enhances the
energy of the desired oscillator, while running the algorithm backwards
spreads out the energy of the perturbed oscillator among its partners.
The former (efficient focusing of energy into a specific oscillator) can
have interesting applications in processes that need crossing of an energy
threshold for completion, and can be useful in nanotechnological devices
and catalysis. The latter (efficient redistribution of energy) can be useful
in processes requiring rapid dissipation of energy, such as shock-absorbers
and vibrational shielding. I present some tentative proposals.
\end{abstract}
\maketitle

\section{Grover's Algorithm}

Database search is an elementary computational task with wide-ranging
applications. Its efficiency is measured in terms of the number of
queries one has to make to the database in order to find the desired item.
In the conventional formulation of the problem, the query is a binary oracle
(i.e. a Yes/No question). For an unsorted database of $N$ items, starting
from an unbiased state and using classical Boolean logic, one requires on
the average $Q=O(N)$ queries to locate the desired item.

Lov Grover discovered a search algorithm that, using superposition of states,
reduces the number of required queries to $Q=O(\sqrt{N})$ \cite{grover}.
The algorithm was originally proposed in the context of quantum computation,
but its applicability has since been widely expanded by realising that the
algorithm is an amplitude amplification process that can be executed by a
coupled set of wave modes. It has also been proved that the algorithm is
optimal for unsorted database search \cite{zalka}.

Grover's algorithm starts with a superposition state, where each item has
an equal probability to get picked, and evolves it to a target state where
only the desired item can get picked. Following Dirac's notation, the
starting and target state satisfy (index $i$ labels the items),
\begin{equation}
|\langle i|s \rangle|^2 = 1/N ~,~~ |\langle i|t \rangle|^2 = \delta_{it} ~.
\end{equation}
The algorithm evolves $|s\rangle$ towards $|t\rangle$, by discrete rotations
in the two-dimensional space formed by $|s\rangle$ and $|t\rangle$. The
rotations are performed as an alternating sequence of the two reflection
operators,
\begin{equation}
U_t = 1 - 2|t\rangle\langle t| ~,~~ U_s = 1 - 2|s\rangle\langle s| ~,
\end{equation}
\begin{equation}
(-U_sU_t)^Q |s\rangle = |t\rangle ~.
\end{equation}
$U_t$ is the binary oracle which flips the sign of the target state amplitude,
while $-U_s$ performs the reflection-in-the-average operation. Solution to
Eq.(3) determines the number of queries as
\begin{equation}
(2Q+1) \sin^{-1} (1/\sqrt{N}) = \pi/2 ~.
\end{equation}
(In practice, $Q$ must be an integer, while Eq.(4) may not have an integer
solution. In such cases, the algorithm is stopped when the state has evolved
sufficiently close to, although not exactly equal to, $|t\rangle$. Then one
finds the desired item with a high probability.)

The steps of the algorithm for the simplest case, $Q=1$ and $N=4$, are
illustrated in Fig.1.

\begin{figure}[t]
\setlength{\unitlength}{0.5mm}
\begin{picture}(150,120)
  \thicklines
\put(15,111){\makebox(0,0)[bl]{Amplitudes}}
\put(65,111){\makebox(0,0)[bl]{Algorithmic}}
\put(65,105){\makebox(0,0)[bl]{Steps}}
\put(120,111){\makebox(0,0)[bl]{Physical}}
\put(120,105){\makebox(0,0)[bl]{Implementation}}
  \put( 0,87){\makebox(0,0)[bl]{(1)}}
  \put(12,87){\line(1,0){32}}
\put(13,99){\line(1,0){2}} \put(17,99){\line(1,0){2}} \put(21,99){\line(1,0){2}}
\put(25,99){\line(1,0){2}} \put(29,99){\line(1,0){2}} \put(33,99){\line(1,0){2}}
\put(37,99){\line(1,0){2}} \put(41,99){\line(1,0){2}}
  \put(45,87){\makebox(0,0)[bl]{0}} \put(45,99){\makebox(0,0)[bl]{0.5}}
  \put(16,87){\line(0,1){12}} \put(24,87){\line(0,1){12}}
  \put(32,87){\line(0,1){12}} \put(40,87){\line(0,1){12}}
  \put(65,93){\makebox(0,0)[bl]{Uniform}}
  \put(65,87){\makebox(0,0)[bl]{distribution}}
  \put(120,93){\makebox(0,0)[bl]{Equilibrium}}
  \put(120,87){\makebox(0,0)[bl]{configuration}}
  \put(28,84){\vector(0,-1){12}}
  \put(30,75){\makebox(0,0)[bl]{$U_t$}}
  \put(65,75){\makebox(0,0)[bl]{Binary oracle}}
  \put(120,75){\makebox(0,0)[bl]{Yes/No query}}
  \put( 0,55){\makebox(0,0)[bl]{(2)}}
  \put(12,55){\line(1,0){32}}
\put(13,61){\line(1,0){2}} \put(17,61){\line(1,0){2}} \put(21,61){\line(1,0){2}}
\put(25,61){\line(1,0){2}} \put(29,61){\line(1,0){2}} \put(33,61){\line(1,0){2}}
\put(37,61){\line(1,0){2}} \put(41,61){\line(1,0){2}}
  \put(45,55){\makebox(0,0)[bl]{0}} \put(45,61){\makebox(0,0)[bl]{0.25}}
  \put(16,55){\line(0,-1){12}} \put(24,55){\line(0,1){12}}
  \put(32,55){\line(0,1){12}} \put(40,55){\line(0,1){12}}
  \put(65,62){\makebox(0,0)[bl]{Amplitude of}}
  \put(65,56){\makebox(0,0)[bl]{desired state}}
  \put(65,50){\makebox(0,0)[bl]{flipped in sign}}
  \put(120,59){\makebox(0,0)[bl]{Sudden}}
  \put(120,53){\makebox(0,0)[bl]{impulse}}
  \put(28,42){\vector(0,-1){12}}
  \put(30,37){\makebox(0,0)[bl]{$-U_s$}}
  \put(65,37){\makebox(0,0)[bl]{Reflection}}
  \put(65,31){\makebox(0,0)[bl]{about average}}
  \put(120,37){\makebox(0,0)[bl]{Overrelaxation}}
  \put( 0,22){\makebox(0,0)[bl]{(3)}}
  \put(12,15){\line(1,0){32}}
\put(13,21){\line(1,0){2}} \put(17,21){\line(1,0){2}} \put(21,21){\line(1,0){2}}
\put(25,21){\line(1,0){2}} \put(29,21){\line(1,0){2}} \put(33,21){\line(1,0){2}}
\put(37,21){\line(1,0){2}} \put(41,21){\line(1,0){2}}
  \put(45,15){\makebox(0,0)[bl]{0}} \put(45,21){\makebox(0,0)[bl]{0.25}}
  \put(16,15){\line(0,1){24}}
  \put(24,15){\circle*{1}} \put(32,15){\circle*{1}} \put(40,15){\circle*{1}}
  \put(65,22){\makebox(0,0)[bl]{Desired state}}
  \put(65,16){\makebox(0,0)[bl]{reached}}
  \put(120,22){\makebox(0,0)[bl]{Opposite end}}
  \put(120,16){\makebox(0,0)[bl]{of oscillation}}
  \put( 0,4){\makebox(0,0)[bl]{(4)}}
  \put(12,4){\makebox(0,0)[bl]{Projection}}
  \put(65,4){\makebox(0,0)[bl]{Algorithm}}
  \put(65,-2){\makebox(0,0)[bl]{is stopped}}
  \put(120,4){\makebox(0,0)[bl]{Measurement}}
\end{picture}
\caption{Amplitude evolution in Grover's algorithm for the simplest case,
$N=4$, when the first item is desired by the oracle. Dashed lines indicate
the average amplitudes.}
\end{figure}
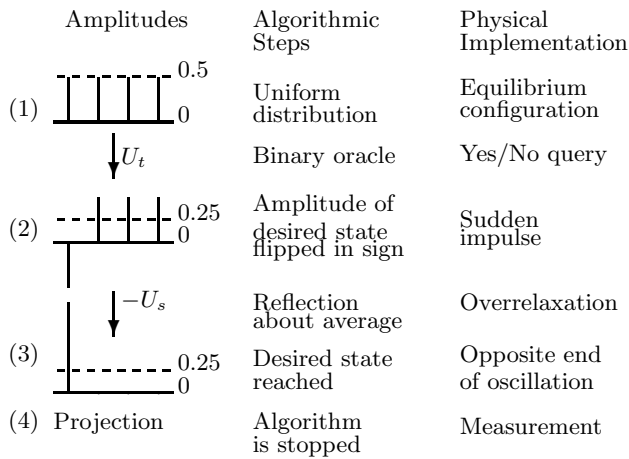

The algorithm relies on superposition of and interference amongst a set
of states, which are generic features of wave dynamics. It can be executed
by any system of coupled wave modes, provided:\\
(1) The superposition of modes maintains phase coherence during evolution.\\
(2) The two reflection operations (phase changes of $\pi$ for the
appropriate mode) can be efficiently implemented.\\
(Note that the $N$ states can be encoded using $\log_2{N}$ bits,
but have to be realised as $N$ distinct wave modes.)

Otherwise, the algorithm is fairly robust, and succeeds even when:\\
(a) The wave modes are anharmonic though symmetric.\\
(b) The initial state is somewhat randomised.\\
(c) The phase changes in the reflection operations are slightly different
from $\pi$,\\
(d) The wave modes are weakly damped.

The interpretation of amplitude amplification occurring in the algorithm
depends on the physical context. In the quantum version, $|A|^2$ gives
the probability of a state, and the algorithm solves the database search
problem. In the classical wave version, $|A|^2$ gives the energy of a mode,
and the algorithm provides the fastest method for energy redistribution
through the phenomenon of beats.

The quantum version of the algorithm involves highly fragile entanglement,
and hence very short coherence times. It also needs to be implemented at the
atomic scale, which is not at all easy. On the other hand, the classical wave
version uses only superposition, which is much more stable, and hence it is
straightforward to design demonstration models \cite{anirvan,wavecatalysis}.
In the following, I describe implementation of Grover's algorithm in a
simple mechanical setting, using four harmonic oscillators coupled via the
centre-of-mass mode.

\section{Harmonic Oscillator Implementation}

A system of coupled harmonic oscillators is frequently studied in physics.
It involves only quadratic forms, and can be solved exactly in both classical
and quantum domains. Let the items in the database be represented by $N$
identical harmonic oscillators. While they are oscillating in a specific
manner, one of them is ``tapped" (i.e. elastically reflected). The task is
to identify which of the oscillators has been tapped, without looking at the
tapping. The optimisation criterion is to design the system of oscillators,
and their initial state, so as to make the identification as quickly as
possible.

\begin{figure}[t]
\epsfxsize=9truecm
\centerline{\epsfbox{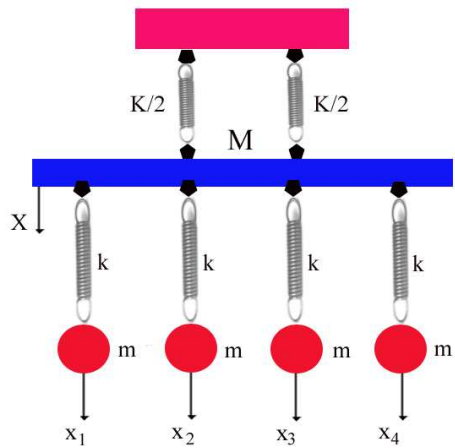}}
\vspace{-2mm}
\caption{A system of four identical harmonic oscillators,
coupled to a big oscillator via the center of mass mode.}
\end{figure}

Grover's algorithm requires identical coupling between any pair of
oscillators. That is arranged by coupling all the $N$ small oscillators
to a big oscillator, as shown in Fig.2. The big oscillator then is coupled
to the centre-of-mass mode, and becomes an intermediary between any pair
of oscillators with the same coupling. Indeed, the centre-of-mass mode
plays the role of the ``average state". In this setting, elastic reflection
of an oscillator implements the binary oracle in velocity space (see Fig.3),
and time evolution of the whole system by half an oscillation period carries
out the reflection about average operation.

\subsection{Dynamics}

The equations of motion are:
\begin{eqnarray}
M\ddot{X} &=& - KX + k\sum_i(x_i - X) ~, \\
m\ddot{x}_i &=& - k(x_i-X) ~.
\end{eqnarray} 
The non-target ($x_{i \ne t}$) oscillators influence the dynamics
of the target oscillator ($x_t$) only through the centre-of-mass
position ${\overline x}\equiv\sum_{i=1}^N x_i/N$.
They make up $N-2$ linearly independent modes, of the form
$x_{j \ne t} - x_{k \ne t}$, which decouple from $x_t$ and $\overline{x}$.
Effective dynamics is thus in the 3-variable space $\{X,\overline{x},x_t\}$,
with the equations of motion:
\begin{eqnarray}
\ddot{X} &=& -\frac{K+Nk}{M}X + \frac{Nk}{M}{\overline x} ~,\\
\ddot{\overline x} &=& -\frac{k}{m}({\overline{x} - X}) ~,\\
\ddot{x}_t - \ddot{\overline{x}} &=& -\frac{k}{m}(x_t - \overline{x}) ~.
\end{eqnarray} 
The last equation is easily solved, with the angular frequency
$\omega_t=\sqrt{k/m}$. The first two equations are coupled,
and their eigenmodes are of the form $X+\lambda{\overline x}$.
We can find them by requiring that
\begin{eqnarray}
\ddot{X}+\lambda\ddot{{\overline x}} &=& X(-\frac{K+Nk}{M}+\lambda\frac{k}{m})
  + {\overline x}(\frac{Nk}{M}-\lambda\frac{k}{m}) \nonumber\\
&=& -\omega^2 (X + \lambda\overline{x}) ~.
\end{eqnarray}
Therefore, 
\begin{equation}
\lambda = \left( \frac{Nk}{M}-\lambda\frac{k}{m} \right)
        \Big/ \left( -\frac{K+Nk}{M}+\lambda\frac{k}{m} \right) ~.
\end{equation}
Let the dimensionless ratios for the spring constants and the masses be,
$R_k=K/k$ and $R_m=M/m$. Then the sinusoidal solutions have the angular
frequency
\begin{equation}
\omega = \omega_t \sqrt{\frac{N+R_k}{R_m}-\lambda} ~.
\end{equation}
There are two solutions to these equations, coefficients $\lambda_\pm$
and the corresponding frequencies $\omega_\pm$. They satisfy
\begin{equation}
\omega_+^2 + \omega_-^2 = \omega_t^2 \left( 1+\frac{N+R_k}{R_m} \right) ~,~
\omega_+^2 \omega_-^2 = \omega_t^4 \frac{R_k}{R_m} ~.
\end{equation}
The general solution to the dynamical equations is:
\begin{eqnarray}
X+\lambda_\pm{\overline x} &=& A_\pm \sin(\omega_\pm t + \phi_\pm) ~,\\
x_t-{\overline x} &=& A_t \sin(\omega_t t + \phi_t) ~.
\end{eqnarray} 

\begin{figure}[t]
\vspace{-5mm}
\epsfxsize=9truecm
\centerline{\epsfbox{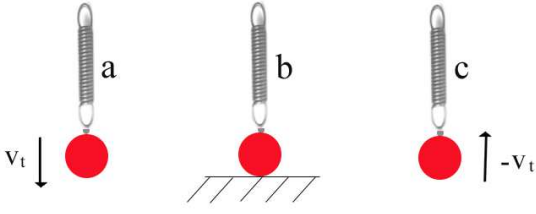}}
\vspace{-5mm}
\caption{The binary oracle flips the sign of the target
oscillator velocity, when its displacement is zero.
(a) Just before the oracle: $x_t=0$, $v_t>0$,
(b) Elastic reflection: $x_t=0$, $v_t \rightarrow -v_t$,
(c) Just after the oracle: $x_t=0$, $v_t<0$.}
\end{figure}

\subsection{The model}

We constructed the simplest system, with $N=4$. The binary oracle (i.e.
$U_t$) is the elastic reflection $\dot{x}_t \rightarrow -\dot{x}_t$ when
$x_t=0$. The reflection about average operation imposes the constraint
that the whole system must evolve by half an oscillation period between
successive oracles. From the many possibilities, as our design parameters,
we selected the convenient frequency ratios
\begin{equation}
\omega_+ = \frac{3}{2}\omega_t ~,~~ \omega_- = \frac{1}{2} \omega_t ~.
\end{equation}
Then the time period $T=4\pi/\omega_t$ for the whole system. Time evolution
for half the period reverses $\dot{\overline{x}}$, while leaving
$\dot{x}_t - \dot{\overline{x}}$ unchanged, i.e. it implements the operator
$U_s$ in the velocity space. Thus Grover's algorithm is realised by ``tapping"
the target oscillator at every time interval $\Delta t = 2\pi/\omega_t$.

The above resonance criterion corresponds to:
\begin{equation}
R_k = \frac{12}{5} ~,~~ R_m = \frac{64}{15} ~,~~
\lambda_+ = -\frac{3}{4} ~,~~ \lambda_- = \frac{5}{4} ~.
\end{equation}
In a situation where all the non-target oscillators move uniformly
together (i.e. all $x_{i \ne t}$ equal, all $\dot{x}_{i \ne t}$ equal),
the displacements are
\begin{eqnarray}
x_t &=& -\frac{1}{2} A_+ \sin(\omega_+ t + \phi_+) +
         \frac{1}{2} A_- \sin(\omega_- t + \phi_-) \nonumber\\
    &+&              A_t \sin(\omega_t t + \phi_t) ~,\\
x_{i \ne t} &=& -\frac{1}{2} A_+ \sin(\omega_+ t + \phi_+) +
                 \frac{1}{2} A_- \sin(\omega_- t + \phi_-) \nonumber\\
            &-&  \frac{1}{3} A_t \sin(\omega_t t + \phi_t) ~.
\end{eqnarray}
Our experimental parameters differed slightly from these ideal values
because of various imperfections discussed later.

\subsection{Results}

The uniform superposition state corresponds to the initial conditions:
\begin{equation}
X=0 ~,~~ \dot{X}=0 ~,~~ x_i=0 ~,~~ \dot{x}_i=V ~.
\end{equation}
In this case, the big oscillator returns to its initial rest state after
every half a period, and the first binary oracle is applied at $t=0$.
The starting phases of the solution vanish, $\phi_\pm=\phi_t=0$. The
amplitudes are
\begin{equation}
A_+ = -\frac{V}{2\omega_t} ~,~~ A_- = \frac{5V}{2\omega_t} ~,~~ A_t = 0 ~,
\end{equation}
before the binary oracle, and
\begin{equation}
A_+ = -\frac{V}{4\omega_t} ~,~~ A_- = \frac{5V}{4\omega_t} ~,~~
A_t = -\frac{3v}{2\omega_t} ~,
\end{equation}
after the binary oracle. The resultant time evolution of the oscillators
in the position and velocity spaces is illustrated in Fig.4. It is observed
that Grover's algorithm provides position amplification of $1.87$ and
velocity amplification of $2$.

\begin{figure} [t]
\epsfxsize=9truecm
\centerline{\epsfbox{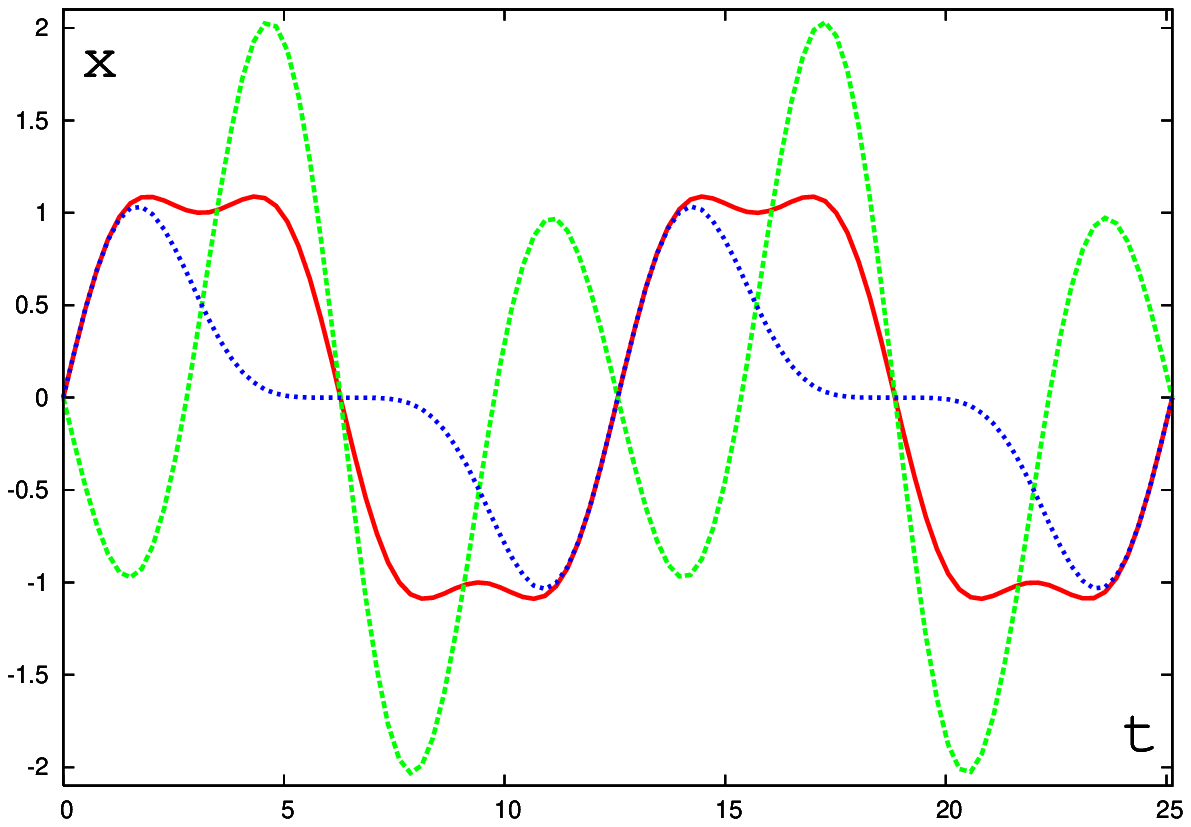}}
\epsfxsize=9truecm
\centerline{\epsfbox{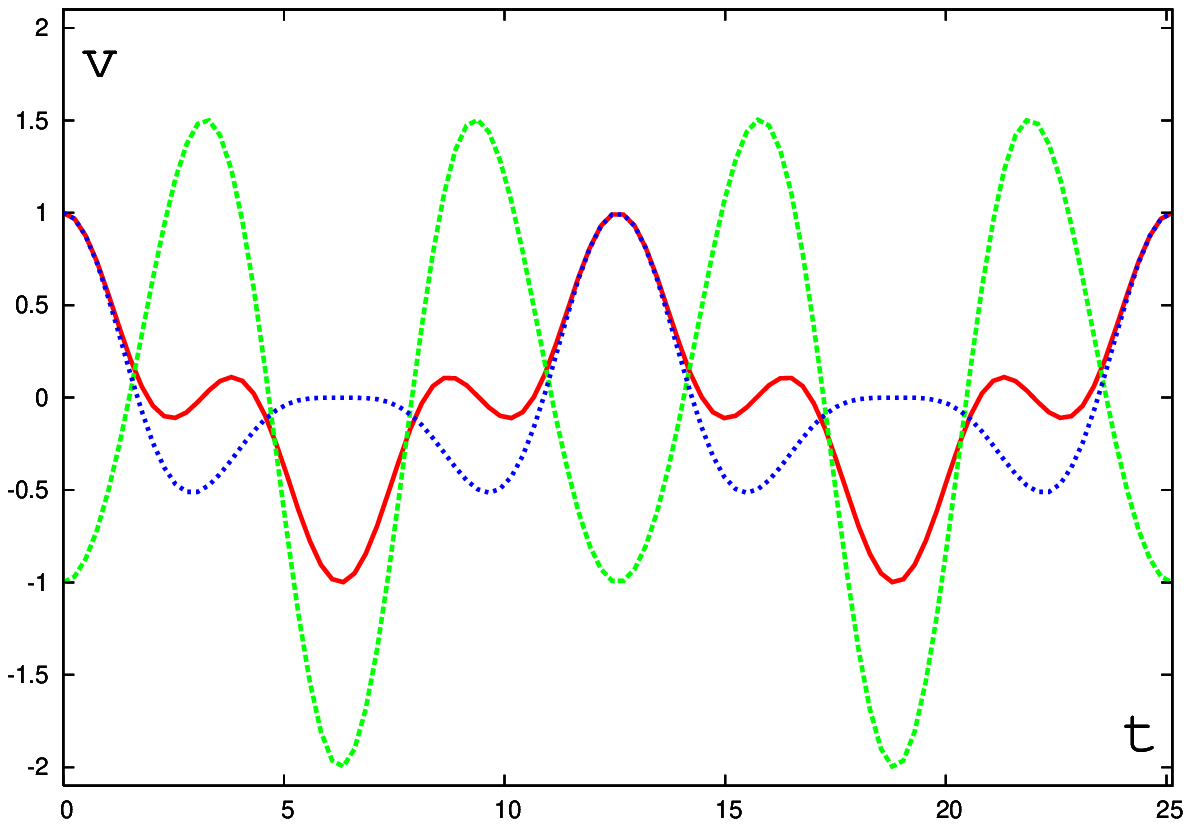}}
\caption{Time evolution of the oscillators in the position (top)
and the velocity (bottom) spaces, when the initial state is:
$X=0,~ \dot{X}=0,~ x_i = 0,~ \dot{x}_i = V.$
Red, green and blue curves respectively denote the uniform superposition
state (without any oracle), the target oscillator after the oracle at
$t=0$, and the non-target oscillators after the oracle at $t=0$.}
\end{figure}

In actual experiment, it is much easier to start with the initial
conditions:
\begin{equation}
X=A ~,~~ \dot{X}=0 ~,~~ x_i=A ~,~~ \dot{x}_i=0 ~.
\end{equation}
In this case, the velocities reach their maximum value after a quarter
period, and the first binary oracle is applied at $t=T/4$. The starting
phases of the solution are, $\phi_\pm=\phi_t=\pi/2$. The amplitudes are
\begin{equation}
A_+ = \frac{A}{4} ~,~~ A_- = \frac{9A}{4} ~,~~ A_t = 0 ~,
\end{equation}
before the binary oracle, and
\begin{equation}
A_+ = \frac{A}{16} ~,~~ A_- = -\frac{21A}{16} ~,~~ A_t=\frac{9A}{8} ~,
\end{equation}
after the binary oracle. The resultant time evolution (as a function of
$t'=t-T/4$) of the oscillators in the position and velocity spaces is
illustrated in Fig.5. It is found that Grover's algorithm provides
position amplification of $1.63$ and velocity amplification of $2$.

In both these cases, maximal velocity amplification is achieved.
On the other hand, position amplification is substantial but not maximal.
The reason is that the binary oracle is implemented in the velocity space,
and it is physically not possible to implement it in the position space.

\begin{figure}[t]
\epsfxsize=9truecm
\centerline{\epsfbox{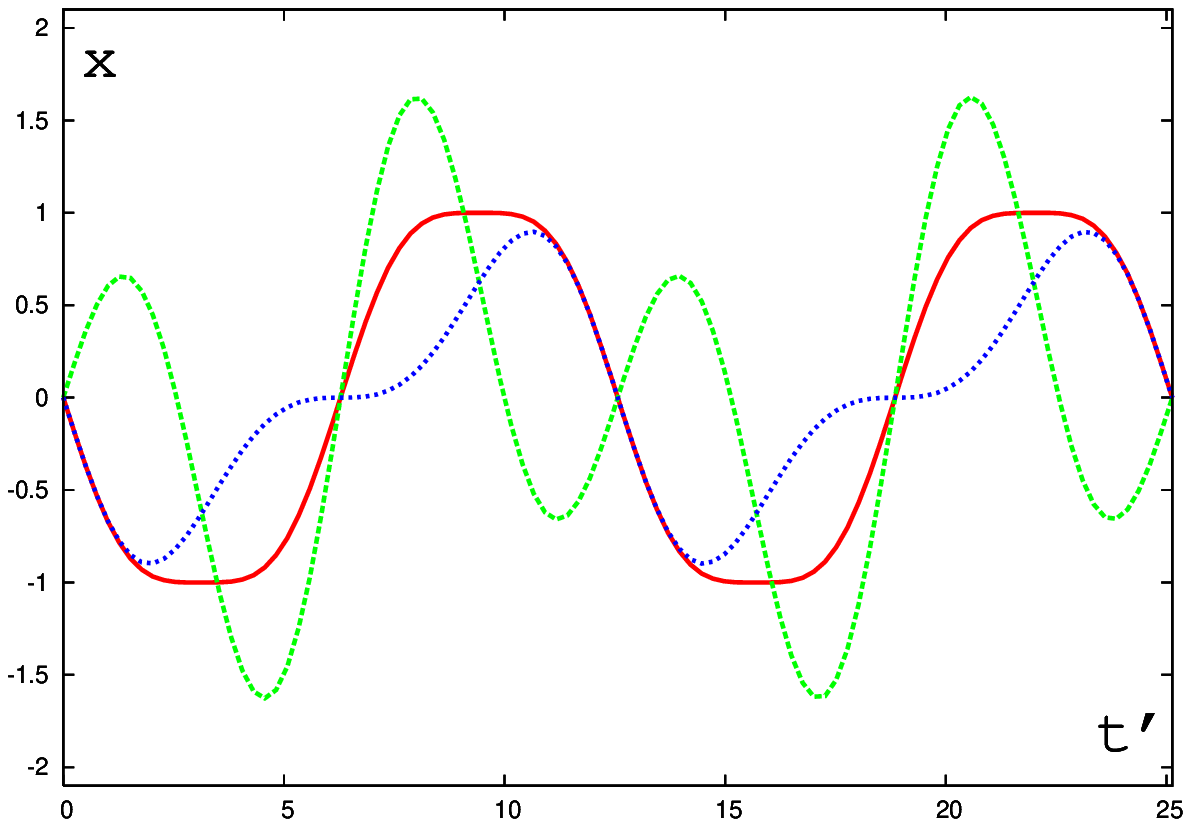}}
\epsfxsize=9truecm
\centerline{\epsfbox{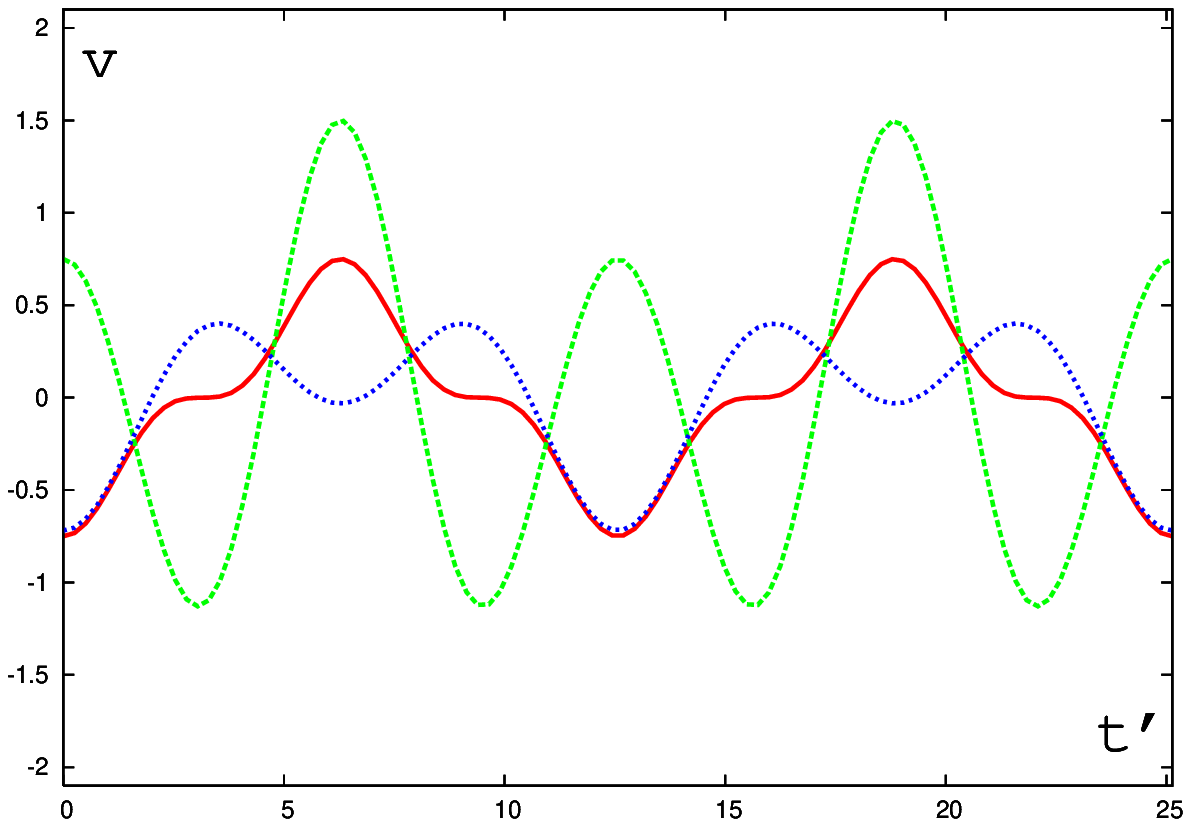}}
\caption{Time evolution of the oscillators in the position (top)
and the velocity (bottom) spaces, when the initial state is:
$X=A,~ \dot{X}=0,~ x_i = A,~ \dot{x}_i = 0.$ Only the behaviour
after the oracle is shown, $t' \equiv t-T/4$.
Red, green and blue curves respectively denote the uniform superposition
state (without any oracle), the target oscillator after the oracle at
$t=T/4$, and the non-target oscillators after the oracle at $t=T/4$.}
\end{figure}

\subsection{Perturbations and stability}

{\it Gravity:}
A uniform gravitational field shifts the equilibrium positions of the
vertically hanging springs, but apart from that it has no effect on their
dynamics. This is obvious from the expression for the potential energy,
\begin{equation}
\frac{1}{2}kx^2 - mgx = \frac{1}{2}k \left( x-\frac{mg}{k} \right)^2
                      - \frac{m^2g^2}{2k} ~.
\end{equation}

{\it Imperfect synchronization of the initial state:}
When the initial velocities are arbitrary instead of uniform, the energy
amplification of the target oscillator provided by the algorithm is not
four-fold. It is instead limited to the initial energy present in the
$\{X,{\overline x},x_t\}$ modes, 
\begin{equation}
\left[ \left( 4\dot{\overline x}^2 +
              \frac{4}{3}(\dot{x}_t - \dot{\overline x})^2 \right)
       \bigg/ \dot{x}_t^2 \right]_{t=0} ~,
\label{maxgain}
\end{equation}
which is still substantial for the generic situation where the initial
$\dot{x}_t$ and $\dot{\overline x}$ do not differ by a large amount.

{\it Imprecise reflection operations:}
In practice, the reflection operations may not exactly implement phase
changes of $\pi$. Also, the measurement operation terminating the algorithm
may not take place at the precise instant of maximum amplification. The
energy amplification depends only quadratically on such phase errors from
the ideal values, e.g. if the reflection phase change is $\pi+\delta$,
then the loss in energy amplification is $O(\sin^2(\delta/2))$.

{\it Inelastic reflections:}
When the binary oracle produces an inelastic reflection of the target
oscillator, with a coefficient of restitution $1-\epsilon$ (i.e.
$v_t \rightarrow (\epsilon-1)v_t$), the energy amplification decreases
by the factor $(1-\epsilon/4)^2$. Even in the extreme case of $\epsilon=1$,
the energy amplification is a sizeable $(9/4)$-fold.

{\it Spring masses:}
Real springs are not massless. To a good approximation, the energy taken
up by the springs can be estimated by assuming a constant velocity gradient
along the springs, and then absorbing the spring energy by altering the
masses of the objects attached to the springs. The dominant correction
is to add one-third of the spring mass to the objects at either end. The
remainder, in our set up and on the average, amounts to adding one-twelfth
of the mass of the small springs to the big mass. This prescription allows
tuning of the masses of the oscillators after measuring the masses of the
springs, in order to maximise the energy amplification.

{\it Damping:}
For a weakly damped oscillator (damping force $-2\gamma mv$), its amplitude
changes linearly with the damping coefficient $\gamma$, while its frequency
changes quadratically \cite{HRK}.
\begin{eqnarray}
&& \ddot{x} + 2\gamma\dot{x} + \omega_0^2 x = 0 ~~\Longrightarrow~~ \nonumber\\
&& x = A e^{-\gamma t} \cos\left(\sqrt{\omega_0^2 - \gamma^2}~t + \phi\right)
\end{eqnarray}
Thus small external disturbances reduce the energy amplification, by a
factor $\exp(-4\pi\gamma/\omega)$, but have little effect on the all
important phase coherence amongst the oscillators that governs interference
of the modes.

Overall, we find that most deviations of the parameters from their ideal
values affect the performance of the algorithm only quadratically, and can
be easily taken care of. Damping provides the only linear perturbation,
which should be controlled to the best possible extent.

\section{Possible Applications}

A variety of coupled vibrational systems with small damping can be made
easily. They can provide either fast focusing or fast dispersal of energy,
and can therefore be an important component in processes sensitive to energy
availability.

\subsection{Focusing of energy}

Efficient concentration of the total energy of a coupled oscillator system
into a specific oscillator can be used as a trigger or a sensor, where an
external disturbance becomes the cause for the reflection oracle.
The focusing of energy could also be useful in nanomechanical systems
\cite{nanotech} where the component concerned cannot be directly controlled,
and a possibility of using coupled cantilever beams as a switch is pointed
out in Fig.6.

\begin{figure}[t]
\epsfxsize=6truecm
\centerline{\epsfbox{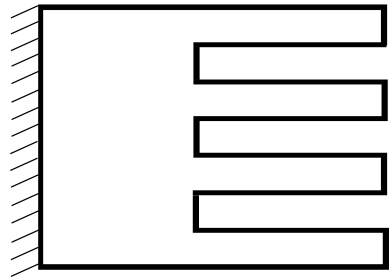}}
\caption{A comb-shaped cantilever beam can be used as a selective switch.}
\end{figure}

There exist many processes that need crossing of an energy threshold for
completion. Their rates are typically governed by the Boltzmann factor for
the energy barrier, $\exp(-E_{\rm barrier}/kT)$. Energy amplification can
speed up the rates of such processes by large factors, leading to catalysis
\cite{wavecatalysis}.

\subsection{Dispersal of energy}

Grover's algorithm is fully reversible. The reflection operators $U_s$ and
$U_t$ are inverses of themselves. So the algorithm can be run backwards as
$ (-U_tU_s)^Q |t\rangle = |s\rangle $. That disperses large initial energy
in the target oscillator to a uniform distribution among its partners. In the
coupled oscillator model, the initial condition would be $\dot{x}_t = V$ and
$\dot{x}_{i \ne t} = 0$. After waiting for $t=2\pi/\omega$, and then reversing
$\dot{x}_t$ produces $\dot{x}_i = V/2$.

This behaviour can be useful in quickly reducing localised perturbations by
redistributing its energy throughout the system. Instead of damping a single
perturbed oscillator, it is much more efficient to disperse the energy into
several oscillators while damping all of them together. To illustrate the
concept, consider the situation where all the oscillators have the same
damping coefficient $\gamma$. The normal modes in the $\{X,\overline{x},x_t\}$
space then separate the same way as in Eq.(10), and the relations in
Eqs.(12,13) are retained. Damping shifts the oscillation frequencies
according to $\omega' = \sqrt{\omega^2 - \gamma^2}$, and the general
solution becomes:
\begin{eqnarray}
X+\lambda_\pm{\overline x} &=& A_\pm e^{-\gamma t}
                               \sin(\omega'_\pm t + \phi_\pm) ~,\\
x_t-{\overline x}          &=& A_t e^{-\gamma t}
                               \sin(\omega'_t t + \phi_t) ~.
\end{eqnarray} 
The coupled oscillator dynamics of Grover's algorithm is maintained by
keeping the frequency ratios unchanged, 
\begin{equation}
\omega'_+ = \frac{3}{2}\omega'_t ~,~~ \omega'_- = \frac{1}{2} \omega'_t ~.
\end{equation}
If the initial conditions are chosen as
\begin{equation}
X=0 ~,~~ \dot{X}=0 ~,~~ x_{i}=0 ~,~~ \dot{x}_t=V ~,~~ x_{i \ne t}=0 ~,
\end{equation}
then after half an oscillation period, $T'/2 = 2\pi/\omega'_t$,
\begin{equation}
\dot{x}_t         =  {1 \over 2}V e^{-2\pi\gamma/\omega'_t} ~,~~
\dot{x}_{i \ne t} = -{1 \over 2}V e^{-2\pi\gamma/\omega'_t} ~.
\end{equation}
This results show that the distribution of energy among the coupled
oscillators suppresses the energy of the target oscillator by an extra
factor of 4, in addition to the usual damping factor for a stand-alone
oscillator. It is indeed the maximum possible reduction in energy,
combining both the mechanisms.

The damping coefficient that maximises the rate of energy loss of the
target oscillator is given by

\begin{equation}
{d \over d\gamma} \left[ {{1 - {1 \over 4}e^{-4\pi\gamma/\omega'_t}}
                         \over {2\pi/\omega'_t}} \right] = 0
~\Longrightarrow~ \gamma = 0.213 \omega_t ~.
\end{equation}
With this choice, in time $T'/2$, the energy of the target oscillator is
reduced to $1.6\%$ of its initial value. Although the energy loss is more
compared to a stand-alone oscillator, it is less localized because it is
distributed among the target oscillator, its partners and the big oscillator.
The corresponding spring constant and mass ratios are:
\begin{equation}
R_k = 2.91 ~,~~ R_m = 4.68 ~.
\end{equation}

A hierarchical system of coupled oscillators (see Fig.7) can be even more
efficient in dispersal of energy, by implementing the above mechanism
simultaneously at multiple scales. The simplest choice would be to couple
four small oscillators to a big one at every level, with appropriate mass,
spring and damping parameters.

\begin{figure}[t]
\epsfxsize=9truecm
\centerline{\epsfbox{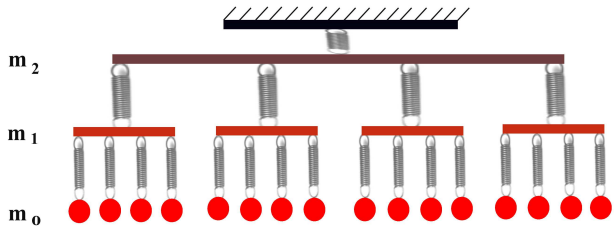}}
\caption{A hierarchical system of coupled oscillators can act as a shock
absorber. The initial impulse is assumed to be a local disturbance, which
subsequently spreads out.}
\end{figure}

\vspace{5mm}
\subsection{Transfer of energy}

It is also possible to combine dispersal and concentration operations to
to transfer energy from one oscillator to another via the centre-of-mass
mode. For example, in case of four coupled oscillators, initial energy in
oscillator $|t_1\rangle$ can be transferred to oscillator $|t_2\rangle$ by
\begin{equation}
(U_sU_{t_2}) (U_{t_1}U_s) |t_1\rangle = |t_2\rangle ~.
\end{equation}
In this manner, a local signal received by a large detector can be
first dispersed over the whole system and then extracted at a specific
location.

\end{document}